# MAS for video objects segmentation and tracking based on active contours and SURF descriptor

Mohamed Chakroun [1], Ali Wali[2] and Adel M. Alimi[3]

[1,2,3] REGIM: REsearch Groups on Intelligent Machines, National Engineering School of Sfax (ENIS), University of Sfax, Sfax, Tunisia

**Abstract**
In computer vision, video segmentation and tracking is an important challenging issue. In this paper, we describe a new video sequences segmentation and tracking algorithm based on MAS "multi-agent systems" and SURF "Speeded Up Robust Features". Our approach consists in modelling a multi-agent system for segmenting the first image from a video sequence and tracking objects in the video sequences. The used agents are supervisor and explorator agents, they are communicating between them and they inspire in their behavior from active contours approaches. The tracking of objects is based on SURF descriptors "Speed Up Robust Features". We used the DIMA platform and "API Ateji PX" (an extension of the Java language to facilitate parallel programming on heterogeneous architectures) to implement this algorithm. The experimental results indicate that the proposed algorithm is more robust and faster than previous approaches.

## 1. Introduction

Video segmentation and tracking is an important process in computer vision because it is the first step of the image understanding process, feature extraction, classification, recognition, and all others steps depend strongly on its results.

In recent years, multi-agent concepts have been used in several areas and they have demonstrated their efficiencies to solve distributed problems. Video processing is one of these complex problems, where multiple applications can be disposed such as segmentation and tracking of moving objects.

The goal of this paper is modelling a video segmentation and object tracking system based MAS "multi agent system" and SURF "Speeded Up Robust Features". The agents used in our MAS "multi agent system" are communicating and inspired in their conduct from active contours approach. They demonstrated their effectiveness to give good results both for segmentation and tracking moving objects.

This paper is organized into four main sections. In the first we present segmentation and object tracking in a video state of the art, in the second, we present active contours approach, in the third section, we present multi-agents paradigm and their applications in image processing, and in the fourth section we describe our proposed algorithm and we present the obtained results of segmentation and objects tracking in a video.

## 2. Segmentation systems: a brief review

MAS "Multi-agent systems" are used in several applications to know segmentation which remains a central task in image processing:

In [1], M chakroun et al presents a Moving Object Segmentation and Tracking based MAS. The agents used are all initialised in the extremity of the video frames and they move to the objects.

Inspirating from biology, the behavior of social spiders is operated through an ADM in order to find a generic approach for image segmentation [2].

J. Fleury [3] proposes a segmentation method using multi-agent systems for multi-object detection, semi-interactive and generic nature, applied to the extraction of cardiac structures in CT imaging.

G. Germond et al. [4] propose a combination of several types of information and knowledge extracted from a multi-agent system, a deformable model and an edge detector to perform the task of segmentation of MRI images. Indeed, the contour is detected by "agents Contours using the method of active contours.

Ouadfel [5] uses "Ant Colony Optimization" to model an algorithm for image segmentation based on MRF (MRF: Markov Random Fields), known as ACS-MRF (AntClust Segmentation-MarkovRandom Fields). In fact, the ACS-MRF is a distributed approach based on ants population considered as reactive agents. Each agent (ant) constructed a set of possible solutions using the pheromone information accumulated by the other ants in a common matrix called pheromone matrix.

## 3. SURF: "Speeded Up Robust Features"

Speeded Up Robust Features (SURF) is used to detect interest points of the image and feature descriptor. SURF is used, in computer vision, for object detection and 3D reconstruction. SURF is partially inspired by SIFT "Scale-invariant feature transform", SURF is faster and more robust than SIFT for different image transformations [15]. SURF is based on sum of Haar wavelet responses and makes effective use of 2D image integrals. SURF uses an approximation of the Haar wavelet blob detector based Hessian determinant.

## 4. Active contours

Active contour models (or snakes) are used in image segmentation and understanding, they are also appropriate for analysis of dynamic image data and they are a framework for delineating an object outline from a possibly 2D image.

Active contour algorithm consists to progress an initial closed contour to an equilibrium position.

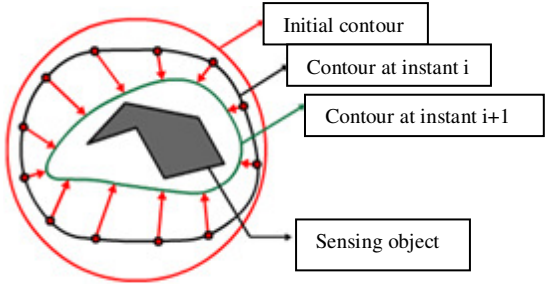

Fig. 1. The principle of active contours

The contour is moving in the direction of the object to be detected. This framework attempts to minimize an energy associated to the current contour as a sum of an internal and external energy:

The energy function is:

$$E_{snake} = E_{int}(v) + E_{ext}(v) \quad (1)$$

- *The internal energy*:

$E_{int}(v)$ is supposed to be minimal when the snake has a shape which is supposed to be relevant considering the shape of the sought object. The most straightforward approach grants high energy to elongated contours (elastic force) and to bended/high curvature contours (rigid force), considering the shape should be as regular and smooth as possible.

$$E_{int}(v) = \int_0^1 (\frac{\alpha}{2}\|v'(s)\|^2 + \frac{\beta}{2}\|v''(s)\|^2) ds$$

$v(s) = (x(s), y(s))$ ($s \in [0, 1]$) is the current point of the contour C and $x(s), y(s)$ are x,y co-ordinates along the contour.

$\alpha$ and $\beta$ are specify elasticity and stiffness of the snake.

$v'(s)$ and $v''(s)$ are the first and second derivatives of v with respect to s.

- *The external energy:*

$E_{ext}(v)$ is supposed to be minimal when the snake is at the object boundary position. The most straightforward approach consists in giving low values when the regularized gradient around the contour position reaches its peak value.

$$E_{ext}(v) = -\lambda \int_0^1 \|\nabla I(v(s))\|^2 ds \quad (2)$$

$\nabla I(v(s))$ is the gradient of the image I in $v(s)$.

$$E_{ext}(v) = -\lambda \int_0^1 \|\nabla(g_0 * I)(v(s))\|^2 ds \quad (3)$$

$g_0$ is gaussien centrée d'écart type $\sigma$.

In Snakes, we use the technique of matching a deformable model to an image by means of energy minimization. A snake initialized near the target gets refined iteratively and is attracted towards the salient contour. A snake in the image can be represented as a set of n points.

In literature, we distinguish two broad categories of segmentation approach by active contour: contour-based approaches [6, 7, 8] and those based on regions [9, 10]. Originally, active contours are contour-based methods. In these approaches, the active contour evolves to the strongest gradients of the image with a partial differential equation that may or may not be deduced from a function including only term based contour. In our case, we are interested in approach-based active contour edges.

O. Michailovich et al. [11] propose a robust method of segmentation using active contours (more precisely the level sets), whose evolution is governed by the flow gradient resulting from an energy function which is based on the Bhattacharyya distance. The results illustrate the effectiveness of this method.

## 5. Proposed System

Our system has two steps: first step is segmentation and the second is the tracking of the video sequence.

### 5.1 Segmentation step

In this step, we use multi-agent system, inspired in their conduct from active contour, and SURF descriptor.

The application of SMA for segmentation consists on:

- Define a society of agents.

- Identify possible actions between agents.

- Define the possible interactions between agents.

Agents used in our system are adaptive type. They are based on the behavior of active contours directed by internal force: $E_{int}(v)$ and external force: $E_{ext}(v)$ (defined in paragraph 4).

The environment of the agents used is the 2D image to be segmented. Once positioned in the image, each agent explores the region in two steps:

- Analyze the local area.

- Consider its neighbors.

The choice of the number and initial position of the used agents in the segmentation system is difficult, in particular if we haven't preliminary information on the type of images to be processed.

**System overview:**

The principle is that at any moment, a pixel may belong to one or more agents. An agent is a pixel. Each agent operates in the image by changing its initial pixel. Each pixel of the image can be processed several times, but we are not obliged to process all the pixels of the image (some pixels at the end, are considered unimportant data). The evolution of our model is based on cooperation among all the explorer agents of the images, which has limited capacity and a partially known environment and are capable of evolution and communication of the supervisor agent with the explorer agent. Each explorer agent is thus led to examine its local neighbors and use its observations. It changes to reach the edges of the / these object (s) to extract. It must also adapt its behavior to consider the presence of other agents. Indeed, each agent regulates the direction of its evolution based on the information received from neighboring agents and the evolution follows the algorithm developed later.

Our system can be seen through two levels, global level and local level as follows:

*- Global level:*

This is the level seen so comprehensive that we note the presence of a "supervisor agent", whose role is to supervise all the agents of the system during their evolution. The supervisor agent decides at some point, if the contour processing an object must be divided or not. The division of stroke involves an increase of objects in the image. A contour is divided when two or more agents are on the same pixel to be processed. Every agent has a knowledge that helps the others to make decisions about all other agents. Once the outline is split, the supervisor agent knows that the membership of each agent has a specific contour.

*- Local level:*

Composed of a set of "explorator agents" This is a level system where agents interact and communicate. They perceive their environment, their local knowledge to help make decisions. An agent at this level has knowledge of the image itself (color, gradient). It can also take decisions as well as for travel agents and their neighbors to move. Collective works are supervised by the supervisor agent and all local agents to detect the contour.

**Agent initialization algorithm:**

Initially, we define the interest points on the first video frame to be segmented using the descriptor SURF "SpeedEDIT Up Robust Features". After, we select eight interest points closest to the extremity of the image. We set eight Explorer agents in the eight feature points.

Step 1: We use SURF "SpeedEDIT Up Robust Features" to define the interest points (figure 2).

Step 2: Selection of the eight closest points in the extremity of the image.

Step 3: Agent initialization: each agent must be positioned on one of the interest points selected in step 2.

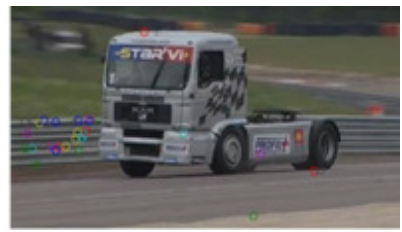

Fig. 2. Define the interest points of the image.

**Interaction between agents:**

The interaction between agents through communication and the determination of the organization of the system.

– Communication between explorer agents and supervisor agent:

It is a communication between the supervisor agent and other agents of the explorer system. The interest of this communication is to manage situations of conflict between staff (case of simultaneous treatment of the same pixel).Through this communication, the agent supervisor can decide if a contour will be divided or not. This communication may be seen as being cooperative.

– Communication between explorer agents:

This is a communication between all agents of the explorer system. The purpose of this communication is to manage the evolution of an agent according to information received from other agents. Indeed, each agent receives an explorer of other agents, their colors and their positions in the image end to determine the direction of evolution specifically. This communication may be seen as being coordinated.

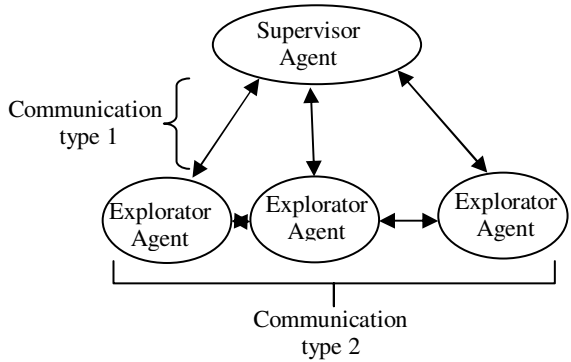

Fig. 3. Interaction between agents

**Description of segmentation algorithm:**

Each agent is initialised on one interest point. It receives eight pixels around it. The number of agents is initially set. There is only one supervisor agent in the system. It begins with the activation of the supervisor agent, scattering explorer agents in the image processed and activate them. The explorator agents move following the active contour method, it consist to minimize the energy E of the contour formed by the connection between all agents:

$$E = E_{int}(v) + E_{ext}(v).$$

### 5.2 Object tracking step

We use the proposed segmentation method, described in the previous paragraph, to segment the first frame of a video sequence and then extract the objects to follow. The contour obtained will serve as a reference for the images that follow.

Algorithm is as follows:

Step 1: Retrieve the positions of the eight interest points selected in the previous frame.

Step 2: Track the motion of the eight interest points in the novel frame, we use SURF for that.

Step 3: Perform the segmentation by taking into account frame i+1 the initial positions of agents in the frame i.

## 6. Experimental results

We tested our segmentation and tracking algorithm on different road traffic videos essentially (see figure 4). The gradient image is the image resulting from the application of the operation "Gradient Magnitude" which is an edge detector that calculates the magnitude of the gradient vector of the image in two orthogonal directions.

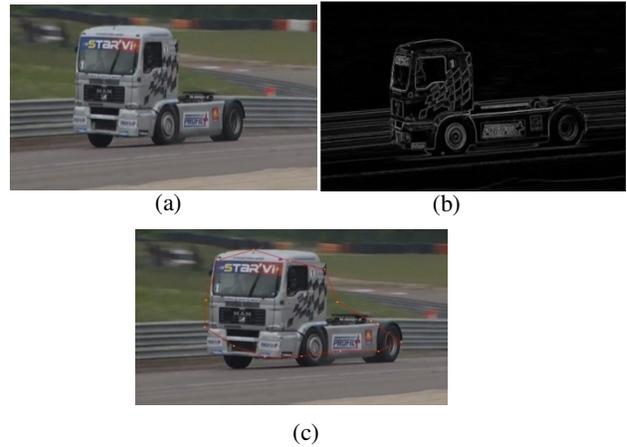

(a) Original image, (b) gradient magnitude of the image, (c) contour result

Fig. 4 . Image segmentation

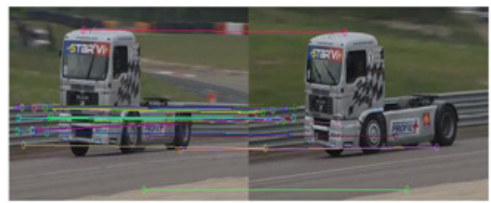

Frame i        Frame i+1

Fig. 5. Tracking based on interest points

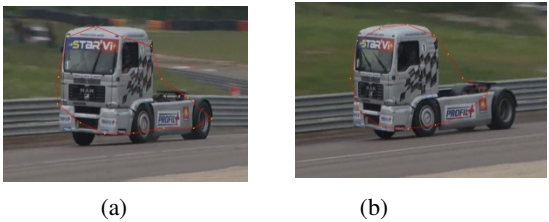

(a)            (b)

Fig. 6. Contour of Frame i (a) and Frame i+1 (b)

## 7. Discussion

To compare our method with methods that use only "Level Sets", we did test the same images used to evaluate our system. We use 20 sequences of video road traffic of 30 minutes each one.

In order to position ourselves versus the methods that use only "Level Sets", we made tests on the same images used to evaluate our system. The following figures show the results of image segmentation. Our system was developed on a multi-core i5 Based on API Ateji PX which is an extension of the Java language to facilitate parallel programming on heterogeneous architectures (multi-core processors, GPUs, clusters).

The performance of our system in terms of accuracy of segmentation as well as execution time.

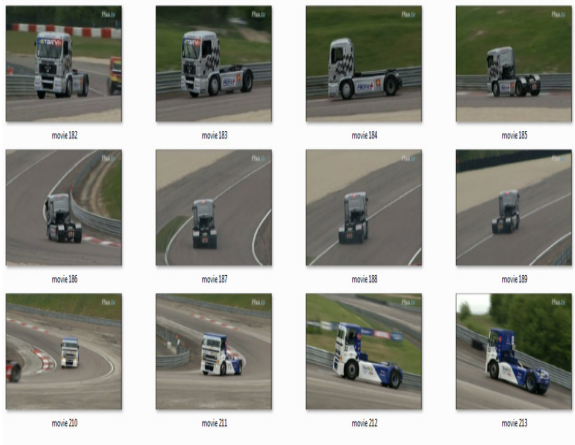

Fig. 7. Sequences of video road traffic

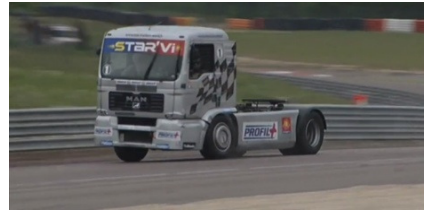

(a)    Original image

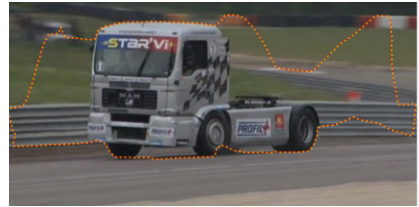

(b)    segmentation with Level Sets algorithm

Fig. 8. "Level Sets" segmentation result (300 iteration)

## 8. Conclusion

We proposed in this paper a new Multi-agent based system to segment and tracking objects based SURF descriptor and Level-set techniques. We have defined the environment of our system, the type and attributes of the agents used and the tasks they must perform. We have implemented our model using the DIMA platform. We have proposed a multi-agent approach for the segmentation of two-dimensional images, the interaction between agents allows detection of objects in the image. Moreover, the choice of the initial distribution of agents in terms of position is based on SURF descriptor. However, our approach remains open to several short-term improvements such as the extension of this process on different areas of images.

### Acknowledgements

The authors would like to acknowledge the financial support of this work by grants from General Direction of Scientific Research (DGRST), Tunisia, under the ARUB program.

**Mohamed Chakroun** (IEEE member), received the B.S. degree in computer science from The Faculty of Sciences of Sfax-Tunisia (FSS) in 2003, and the Master degrees in Computer Science from the National Engineering School of Sfax - Tunisia (ENIS), in 2006. In September 2006, he joined the Sfax University (USS), where he was an assistant professor in Preparatory Institute for Engineering Studies of Sfax (IPEIS).

**Ali Wali** (Phd. Eng.), received the B.S. degree in computer science from The National Engineering of Sfax-Tunisia (ENIS) in 2003, and the Master degrees in Computer Science from the National Engineering School of Sfax -unisia (ENIS), in 2005. He obtained a PhD in Computer Engineering in 2013 From Sfax University, Tunisia. In September 2005, he joined the Sfax University (USS), where he was an assistant professor in the Department of Computer science of the higher Institute of computer science and multimedia of sfax (ISIMS). He is member of the REsearch Group on Intelligent Machines (REGIM). His research interests include Computer Vision and Image and video analysis. These research activities are centered around Video Events Detection and Pattern Recognition. He is a Graduate Student member of IEEE. He was the member of the organization committee of the International Conference on Machine Intelligence ACIDCA-ICMI2005, Third IEEE International Conference on Next Generation Networks and Services NGNS2011 and 4th International Conference on Logistics LOGISTIQUA2011.

**Adel M. Alimi** (S'91, M'96, SM'00). He graduated in Electrical Engineering in 1990. He obtained a PhD and then an HDR both in Electrical & Computer Engineering in 1995 and 2000 respectively. He is full Professor in Electrical Engineering at the University of Sfax since 2006. Prof. Alimi is founder and director of the REGIM-Lab. on intelligent Machines. He published more than 300 papers in international indexed journals and conferences, and 20 chapters in edited scientific books. His research interests include applications of intelligent methods (neural networks, fuzzy logic, evolutionary algorithms) to pattern recognition, robotic systems, vision systems, and industrial processes. He focuses his research on intelligent pattern recognition, learning, analysis and intelligent control of large scale complex systems. He was the advisor of 24 Ph.D. thesis. He is the holder of 15 Tunisian patents. He managed funds for 16 international scientific projects. Prof. Alimi served as associate editor and member of the editorial board of many international scientific journals (e.g. IEEE Trans. Fuzzy Systems, Pattern Recognition Letters,NeuroComputing, Neural Processing Letters, International Journal of Image and Graphics, Neural Computing and Applications, International Journal of Robotics and Automation, International Journal of Systems Science, etc.). He was guest editor of several special issues of international journals (e.g. Fuzzy Sets & Systems, Soft Computing, Journal of Decision Systems, Integrated Computer Aided Engineering, Systems Analysis Modelling and Simulations). He organized many International Conferences ISI'12, NGNS'11, ROBOCOMP'11&10, LOGISTIQUA'11, ACIDCA-ICMI'05, SCS'04ACIDCA'2000. Prof. Alimi has been awarded with the IEEE Outstanding Branch Counselor Award for the IEEE ENIS Student Branch in 2011, with the Tunisian Presidency Award for Scientific Research and Technology in 2010, with the IEEE Certificate Appreciation for contributions as Chair of the Tunisia Computational Intelligence Society Chapter in 2010 and 2009, with the IEEE Certificate of Appreciation for contributions as Chair of the Tunisia Aerospace and Electronic Systems Society Chapter in 2009, with the IEEE Certificate of Appreciation for contributions as Chair of the Tunisia Systems, Man, and Cybernetics Society Chapter in 2009, with the IEEE Outstanding Award for the establishment project of the Tunisia Section in 2008, with the International Neural Network Society (INNS) Certificate of Recognition for contribution on Neural Networks in 2008, with the Tunisian National Order of Merit, at the title of the Education and Science Sector in 2006, with the IEEE Certificate of Appreciation and Recognition of contribution towards establishing IEEE Tunisia Section in 2001 and 2000. He is the Founder and Chair of many IEEE Chapters in Tunisia section. He is IEEE CIS ECTC Education TF Chair (since 2011), IEEE Sfax Subsection Chair (since 2011), IEEE Systems, Man, and Cybernetics Society Tunisia Chapter Chair (since 2011), IEEE Computer Society Tunisia Chapter Chair (since 2010), IEEE ENIS Student Branch Counselor (since 2010), He served also as Expert evaluator for the European Agency for Research. since 2009.